# Platial mobility: expanding place and mobility in GIS via platio-temporal representations and the mobilities paradigm

Farrukh Chishtie[1,2] · Rizwan Bulbul[3] · Panka Babukova[3] · Johannes Scholz[3]

**Abstract**
While platial representations are being developed for sedentary entities, a parallel and useful endeavor would be to consider time in so-called "platio-temporal" representations that would also expand notions of mobility in GIScience, that are solely dependent on Euclidean space and time. Besides enhancing such aspects of place and mobility via spatio-temporal, we also include human aspects of these representations via considerations of the sociological notions of mobility via the mobilities paradigm that can systematically introduce representation of both platial information along with mobilities associated with 'moving places.' We condense these aspects into 'platial mobility,' a novel conceptual framework, as an integration in GIScience and the mobilities paradigm in sociology, that denotes movement of places in our platio-temporal and sociology-based representations. As illustrative cases for further study using platial mobility as a framework, we explore its benefits and methodological aspects toward developing better understanding for disaster management, disaster risk reduction and pandemics. We then discuss some of the illustrative use cases to clarify the concept of platial mobility and its application prospects in the areas of disaster management, disaster risk reduction and pandemics. These use cases, which include flood events and the ongoing COVID-19 pandemic, have led to displaced and restricted communities having to change practices and places, which would be particularly amenable to the conceptual framework developed in our work.

**Keywords** Place in GIScience · Platial modeling · Platial mobility · Platial displacements · Platio-temporal' mobilities · Platial mobility for disaster management

**JEL Classification** C60 · D80 · H84 · J60 · L86

---

Farrukh Chishtie and Rizwan Bulbul have contributed equally to this work.

✉ Rizwan Bulbul
bulbul@tugraz.at

Extended author information available on the last page of the article

## 1 Introduction

The relevance and importance of 'place' is becoming increasingly important in Geographic Information Science (GIScience) and related disciplines. Goodchild (2011) raised key concerns about existing Geographic Information systems (GISs) driven by technologies that favor the implementation of a strict notion of space. Space is mostly defined in terms of Euclidean geometry and the Cartesian coordinate systems, and it is contrasted with place in the literature because the latter is conceptualized as constructed by humans. Based on both philosophical and humanist notions, there are many representations of how platial information be integrated in existing GISs, that are presently based on 'space'.

In this work, we aim to present a formalism for platial representations that is time dependent and does not consider 'space' and 'place' to be necessarily separated. Moreover, in the consideration of time and movement, we introduce the idea of 'platial mobility' that draws upon the mobilities paradigm (see, e.g., Hannam et al. 2006 and Urry 2016) and is a generalization of various notions of place in the literature that are 'static' in their definition. In other words, by introducing such mobilities via sociological considerations in platial representations, we are also generalizing the usual space-dependent notion of mobility toward including further consideration of place and time in GISs.

The paper is organized as follows. We begin with the introduction to the distinctions within the notion of spaces and places. In the following section, we introduce the notion of 'moving places' and its platial representations, working with observables which have the possibilities of having both space and places connected in relationships. Since places and their movements are humanly and socially constructed, we then introduce the key ideas of the mobilities paradigm in sociology, and how these can help generalize the notions of mobility within GIScience which are Euclidean space and time-dependent. The resulting concept, "platial mobility", is then outlined. In these sections, we provide a thorough literature review of place-based GIS and mobility along with how these notions are used in other disciplines.

With the platial mobility formalism defined, we next discuss the means through which various platio-temporal representations and platial mobility can be formally identified and studied is explored in various cases. This includes the use cases of the Lake Attabad[1] disaster which occurred in 2012 and led to the movement and displacements of a vast section of the population, the relocation in the Eferding Basin, caused by severe flooding events in 2002 and 2013,[2] and the ongoing COVID-19 pandemic. We then discuss how our ideas of "platial mobilities" can be useful in better representing such complex phenomena through a fundamental and joint consideration of both spaces and places, as well as accounting for the sociological aspects in such large-scale movements and disasters.

---

[1] https://en.wikipedia.org/wiki/Attabad_Lake.
[2] https://climate-adapt.eea.europa.eu/metadata/case-studies/relocation-as-adaptation-to-flooding-in-the-eferdinger-becken-austria.

Overall, we consider the various implications of our work and how it can contribute to existing developments of platial representations in GIScience. Our approach is based on pragmatic and practical considerations of how to formalize and generalize platial representations using both mathematical/geometric and sociological considerations. We contend that the platial mobility framework is a paradigm shift for GIScience with new ways to collect and integrate spatio-temporal data with qualitative data to provide a holistic means for both developing better understandings and usefulness for end users and public. This framework has potential for improving existing approaches and its relevance could be significant especially in an era of increasing global-scale crises such as climate change impacts and pandemics.

## 2 The platial turn—comparison of spaces and places

According to Tuan (1977), place is a space infused with human meaning. And this meaning comes with human experiences, which according to Jordan et al. (1998) involves three subjective components called perception, cognition, and the affection. The concept of place needs to therefore accommodate subjective views of individuals, their activities as well as environment in which they act. Blaschke et al. (2018) consider the notion of place that takes form as a result of spatial experiences. Thus, places are subset of meanings mapped by humans to make sense of their environment which are typically located in space and time. In this sense, places are semantically constructed spaces or set of spaces that have spatial (geometry), temporal (time) and meaning (semantics) dimensions and the optional level of detail as a dimension if the places are constructed hierarchically. Other researchers have sought to provide alternative representations. Janowicz et al. (2012) propose geospatial semantics, while McKenzie (2015) conceptualizes place in a spatio-temporal manner while taking subjectivity into account. Gao et al. (2017) use "points of interest (POI)" in a statistical approach to characterize place and its features.

Places may be well bounded having well-defined spatial boundaries, or may also have fuzzy or even dynamic extents (Smith 1995; Scholz et al. 2016). Hence, fuzziness and dynamical features of places indicate that stability of meaning/experience of place (and time) are important considerations and further point to the stability of platial representations themselves.

Places may have hierarchical (place/sub-place) relationships that are valuable in performing platial analysis. The hierarchies are defined along spatial and semantic dimensions thus enabling the level of detail-based representation of places. However, it is not necessary that the two hierarchies match, as the hierarchies in the semantic dimension capture the hierarchical cognitive sense of place that may differ based on an individual or group-based experiences. This makes modeling semantic hierarchies for places quite a challenging task. However, several ontological (Ballatore 2016) and other approaches (Wu et al. 2019) can be used to uncover the semantic hierarchies among vague places.

In order to get a better understanding of the sense of place (Agnew 2011) and contribute toward the formalism of place definition, we need to briefly discuss three important aspects of platial classification, platial construction and platial comparisons. Places can be classified into different types based on some of the salient characteristics that are associated with the spatial extent (or boundary) of the place, stability of the place in time, mobility of the place and last but not least the instantiation (coming into existence) of the place. Table 1 shows the classification of places with a brief description.

Since places are semantically constructed spaces by individuals or groups, that may have sense in three different scales arranged hierarchically from local to global levels as shown in Fig. 1. At the highest level of detail, platial model should be able to represent/handle individual or personal experiences of place. This is challenging from modeling perspective but is much needed for having citizen-centric GIScience (Blaschke et al. 2018) and personal GISs (Abdalla and Frank 2011). Group of individuals or communities may construct places together that may have sense for the group only.

Places can be compared for similarity using the distance measure among places along spatial, temporal and semantic dimensions. How similar are place A and place B along spatial and/or temporal dimensions? Existing spatio-temporal models are sufficient enough to perform similarity measures along spatial and temporal dimensions. Similarly topological relationships between places can be computed using conventional geospatial techniques that utilize platial geometric extents.

However, the computation of similarity between two places along semantic dimension, called semantic similarity, is complex because of inherent subjectivity of the properties defining the semantic meaning of place. Platial semantic similarity can have different contexts. For example, how to compare two cities (places) Paris and Vienna? One option is to consider the cities as the collection of places. In this case, the contextual distribution patterns of places can be analyzed and turned into a similarity measure. Other possibilities are taking into account any contextual theme, e.g., functional diversity and/or availability of places, mobility dynamics, air quality, transportation, disaster resilience, etc. Similarly, two restaurants can be compared based on personal experiences like quality of food, service quality, environment, cleanliness, etc.

Several approaches (e.g., Baikousi et al. 2011; Cox and Cox 2008; Janowicz et al. 2008; Schwering 2008), for modeling similarity measure along multidimensional qualitative data, can be used by coding the semantic platial attributes as dimensions. It is then also possible to assign weights to these dimensions to express relative importance and also combine these with spatial and time dimensions, for example, as shown by Furtado et al. (2016) for measuring multidimensional similarity for semantic trajectories.

## 3 Platial modeling—the formalism of place

Platial representation has got much attention in GIScience more recently. However, the full integration of the concept of place in geospatial tools will only be possible if the notion of place can be rigorously defined and readily formalized (Goodchild 2011) and rests on formal foundations. Looking at the literature reveals that several attempts have already been made in this direction, for example, Jordan et al. (1998);

Table 1 Platial classification

| Type and characteristics | Description |
| --- | --- |
| Dynamic vs Static *Characteristics*: boundary, time | Places can be dynamic or static based on their extents or boundaries. Dynamic places possess flexible boundaries that may change over a time interval, while static places have fixed boundaries that may not change over time. For example, the place of a demonstration may be important for law enforcing agencies for crowd management. However, the extent of the place of demonstration may change continuously over time based on the number of participants and crowd movement dynamics. Static places, on the other hand, have static boundaries that remain invariant over a time interval. For example, the place of a city festival happening in a city hall has static boundaries that will not change substantially over the course of the event |
| Temporary vs Permanent *Characteristics*: lifespan, time | Temporary places have predefined life spans and disappear over time. Permanent places have no predefined lifespans and exist until natural or active human interventions that make these disappear over time. An example of a temporary place is a police check post along a highway that is established for random snap checking. Statue of liberty in USA and Eiffel tower in France are permanent places as these will exist until natural or human interventions are made. The world trade center in USA was a permanent place but (negative) interventions made the place to disappear |
| Movable vs Immovable *Characteristics*: mobility, displacement, time | Movable places are places that may displace or change location over time without losing their essence. These location-based displacements may cause change in other properties as well and can be human induced or natural. For example, a shop or business moved from one place to another. This could be due to human intervention to relocate business to maximize profit or reduce costs, or because of some natural calamity that damaged the place where the business was located and thus making it inevitable to migrate the business to a new place. Immovable places may lose essence (and sense of place) if displaced and are thus immovable. For example, some sacred religious places make sense only being there such as the sacred place of "Kaaba" in the city of Mecca, Saudi Arabia and another such example is the holy "Wailing Wall" in the city of Jerusalem |
| Instantaneous vs Planned *Characteristics*: existence, instantiation, time | Unplanned or instantaneous places come into existence without being planned through the instantaneous occurrence of any event. For example, the place of an accident instantaneously comes into being when a sudden event (the accident) occurs. On the other hand the planned events can create planned places. For example, the planned meeting place, the planned place for constructing a home, etc |

Hockenberry (2006); Goodchild and Li (2011); Roche (2016, 2017); Papadakis et al. (2017); Acedo et al. (2018); Papadakis et al. (2019), and most of the literature provide theoretical underpinnings toward platial modeling and some concrete platial modeling attempts. In the following, we summarize some of the attempts for platial modeling. The list is non-exhaustive and briefs some of the approaches that got our attention.

The work by Agarwal (2005) discusses an ontological model. Although a concrete platial modeling technique is not discussed, rather an approach is expanded on the inherent vagueness associated with the concept of place within and across various domains that could be resolved once a definitive semantic framework for place is defined and relationships to these other related concepts were explored. Some experiments were performed to explore such semantic distinctions among place and other related concepts like region and neighborhood, etc. It was shown that region is semantically closer to place than neighborhood. It was inferred that region is a super-type for place and that place is conceptualised as a subset of region. The results depend on the experimental setups and surely do not generalize.

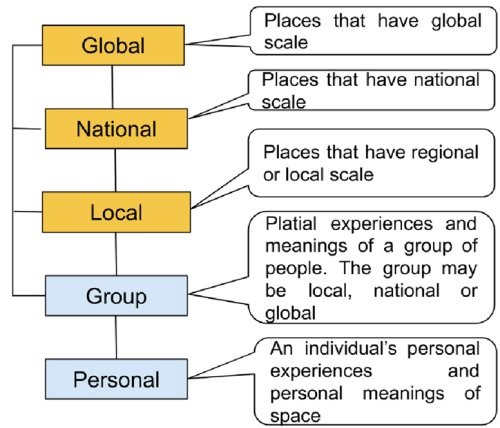

**Fig. 1** Places and scale. Places may have different meanings at different levels based on individual or group experiences

The work by Jordan et al. (1998) discusses an affordance (based) model. The authors discussed the idea of modeling place using the concept of affordances. Affordances are what objects or things offer people to do with them and thus create potential activities for individuals. Affordance of objects depends on individuals' perception and is scale-based. The suggested model uses means-end hierarchy to abstract place. It contains purpose, three functions (abstract, generalized and physical) and the physical form. Affordances at the lower levels of the hierarchy are constrained by affordances at higher level. They suggest the following aspects of place to be used for affordance-based modeling of space: (i) Physical features-Places as collection of objects. (ii) Actions-People perform actions in a place. (iii) Narratives-Storytelling and historical record at all levels. (iv) Symbolic representation-Some places may be referenced by symbols. (v) Socioeconomic and cultural factors-socioeconomic and cultural characteristic of a place afford different behavior in places. (vi) Typologies-People categorize places for dealing with complexity and new situations. Although the approach was discussed in detail, the actual implementation to show the practicality and viability of the model was not shown.

The (functional) model by Papadakis et al. (2017, 2019) provides an ontological model of place based on the notion of functions. The proposed approach is a multi-dimensional model in which dimensions define a place as a set of functions, which entail a spatial structure expressed in patterns of spatial descriptions. Places are defined as spaces infused with functional context that is achieved by converting interrelated components that support certain functions. They have also shown the utility of the approach in place-based searches. The model seems more conducive for categorization of formal places or higher level places. People sometimes play football or different games in a basket ball ground, for example. Strict functional categorization of places is useful but may lead to restrictive definitions by missing the complicated and diversified human experiences.

Many spatial data representation modeling techniques already exist that provide reasonable approximations for the digital representation and manipulation of objects using formal approaches having mathematical foundations. In the absence of such models for the digital representation of places (platial modeling), there is a dire need to design formal platial models that have sound theoretical foundations for platial data representation at multiple levels of detail and support platial operations. Thus, a practical platial model should fulfill following requirements:

1. Supports the platial hierarchical representations at multiple levels of detail.
2. Supports platial and platio-temporal operations.
3. Utilizes the underlying spatial data model for spatial part of place.
4. Considers space, time and meaning (contextual semantics) as compositional dimensions of place.

In order to fulfill the aforementioned requirements, our platial modeling approach defines place as a function dependent on four dimensions.

$$\text{Place} = f(\text{Level}, \text{Time}, \text{Space}, \text{Meaning}) \quad (1)$$

where Level = level of platial detail, Time = time stamp or interval, Space or platial Geometry or Extent = $\cup_{i=1}^{n} geom_i$ (platial extent that captures geometric information; can be implicit as derived from upper level or from corresponding level in spatial model or explicitly defined), Meaning = $(d_1 : v_1, d_2 : v_2, ..., d_n : v_n)$ (a multidimensional structure having $n$ dimensions as key value pairs).

The most distinguishing feature or component of platial definition is the "meaning" aspect as the other components can be handled using existing spatial models. Thus, platial meaning itself is a multidimensional structure that is intentionally segregated though could be treated as normal set of attributes of the platial definition. However, keeping the meaning separate would be useful for performing semantic platial queries that would mostly be relying on qualitative data that are captured in the meaning. The qualitative meaning dimensions can be converted to quantitative dimensions using many of the existing conversion techniques. The converted multidimensional data sets then can be analyzed using well-known multidimensional data analysis techniques, specially from the spatial data mining domain (Jiwai and Kamber 2012; Li et al. 2015). It is not necessary to have fixed set of parameters or a fixed schema for modeling platial meaning. This is practical even for implementation of platial models with NoSQL databases.

In order to understand the relationship between spatial and platial models, Table 2 shows the difference between the two models along the four key dimensions.

## 4 'Platial mobility': considering spatial and time dimensions together

Space and place, and inclusion of time are interconnected concepts in our approach and their formalism is based on pragmatic human geography, while constructions of place without spatio-temporal dependencies are covered using the (mobilities) social science basis/ontologies. McKenzie (2015) has also shown the importance of temporal dimension for platio-temporal research, while Tuan (1977) has extensively discussed the relationship between place and time. Adding temporal dimension would make sense in platial construction and this notion of space-time-based definition of place makes the basis for defining "platial mobility" for extending the notion of purely space-time based notion of "mobility" in GIScience.

A space at a specific time instance or time interval may refer to a different place or places for different persons as shown in Fig. 2. Here, their specific notions and meaning matter distinctly in platial representations. Variability of places, the divergence of two places in meaning, can then be formulated in terms of their spatial (S) and temporal (T) dimensions as shown below.

$$\text{Place}X(S', T', \text{meaning}) = \text{Space\_time}X(S, T) + \epsilon_1(S', T', \text{meaning}) \quad (2)$$

$$\text{Place}Y(S', T', \text{meaning}) = \text{Space\_time}Y(S, T) + \epsilon_2(S', T', \text{meaning}) \quad (3)$$

where $\epsilon$ = variability of space and time from the exact space time location. Variability can introduce statistics, fuzzy logic and other mathematical representations. Deviation between place X and place Y is the difference between the variabilities of specific places. This variability can also be mapped in comparison with exact spatio-temporal location/layers. Since we are discussing place in the context of platio-temporal representations, notions of mobility can also be represented via this framework.

As a consequence to the above definitions, we can further denote related observables $O(PlaceX(S', T', meaning)$ to represent platial representations of distances, speed, etc. For example, with the above formulation, we can not only map platial representation of locations, but distances as well as time taken (or speed) to locations via sampling across a section of a population via interviews, surveys, etc.

Various platial constructions are based on possible combination of space time variations. In order to understand the variations, two possibilities for each space and time dimensions are taken into account. For any place under consideration, the space defining the place remains the same or may change over time. Combination of space time instances is shown in Table 3. For notions of place which are not based on spatial and temporal considerations, we represent using the mobilities paradigm, which is outlined in the next section.

**Table 2** Difference between spatial and platial models

| Dimension | Spatial model | Platial model |
|---|---|---|
| Level of detail | Most of the existing spatial data models support level of detail through representation of spatial data at multiple levels (Frank and Timpf 1994; Davis and Laender 1999; Zhou and Jones 2003) and using hierarchical data structures (Samet 1990; Timpf and Frank 1997; Bulbul and Frank 2009) | In case of platial models, the hierarchical representation of platial information would be the useful option to capture the inherent hierarchical cognitive sense of place. However, it is not necessary that the platial hierarchies match the spatial hierarchies |
| Time | Time is a key dimension for spatio-temporal analysis and can be modeled either as an attribute of n-dimensional geometries (resulting in n.5 D models) or integrated as an additional dimension of the n-dimensional geometries (resulting in n+1 D models) | Time could be a time stamp or interval and is an integral component or dimension of platial definition. Places make sense only in temporal dimension and that also enables platial representation models to support platio-temporal analysis. Time in platial models, for a specific level of detail, is associated with spatial and meaning dimensions |
| Geometry | Many geometric representation data models exist that can handle multi-dimensional geometries. Most of these models can also handle level of detail and support robust geometric and topological operations. A subset of models also support fuzzy and imprecise boundaries | The platial geometries are defined by the space dimension of the place and spatial modeling techniques are sufficient to handle the requirements of platial geometric representations. Platial analysis along spatial dimension utilizes existing spatial analysis techniques |
| Meaning | Semantic enrichment of spatial models is possible using various approaches, e.g., linked data (Kuhn et al. 2014; Iwaniak et al. 2016; Hart and Dolbear 2016) and VGI (Goodchild 2007; Kuhn 2007) | Meaning is the most important and distinguishing dimension of platial model and has the capability to capture cognitive sense of place at various levels of detail. Meaning then makes it possible to perform platial analysis using both geometric and semantic measurements. Like spatial models, semantic enrichment of platial models is also possible. However, this is only viable if the platial dimensions are fixed, well-known and well-controlled (Seamon and Sowers 2008) |

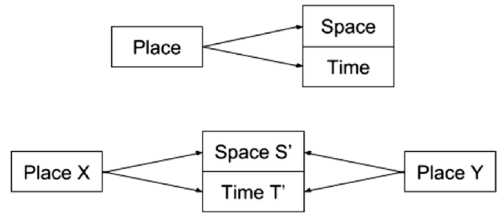

**Fig. 2** Variability of places is a function of space and time dimensions. Two places (*X* and *Y*) may share same space (*S'*) at a given time(*T'*)

## 5 Connecting and enhancing 'platial mobility' with sociology via the mobilities paradigm

By definition, platial representations rely on human based notions, hence sociological considerations are important. To further enhance our notion of "platial mobility" with sociology that includes notions of place which do not have explicit spatio-temporal dependencies, we therefore connect to the mobilities paradigm, a contemporary and rich theoretical and methodological approach in the social sciences.

The mobilities paradigm, based in the social sciences, is a contemporary framework which studies the movement of people, ideas and things, as well as the broader societal consequences of these flows. Urry (2012, 2016) calls for a move to a "borderless" notion of sociology which is not hampered by the confining construct of "society" as a bounded entity, especially in the contemporary times where, for example, globalization and related flows are intensifying.

The "mobility turn" started in the 1990 s, by giving due importance to movements of individuals and groups, basing itself as a critique of sedentarism and deterritorialisation in the social sciences at the time. Since the social sciences traditionally had a static take on the 'social,' the mobility turn raised the significance of travel in everyday life to leisure travel among other types of movements. Sheller and Urry (2006) observe that "sedentarism treats as normal stability, meaning, and *place*, and treats as abnormal distance, change, and placelessness. Sedentarism is often derived loosely from Heidegger (1993), for whom dwelling (or wohnen) means to reside or to stay, to dwell at peace, to be content or at home in a *place*." As implied in these statements, the mobilities paradigm also critiques a static notion of place, which is conceptualized as static by various theorists. Besides the sedentary notion of place by Heidegger (1993), Gieryn (2000), for example, also defines place having three key properties, namely having a "geographic location," "material form" and "investment with meaning."

Beyond such conceptualizations, the mobilities paradigm utilizes novel ways of understanding how movements can better explain formation of identities, networks and everyday life which features greater movement than ever before (Cresswell 2011).

The mobilities research paradigm has taken seriously 'the material turn' and 'the spatial turn' in the social sciences. Influenced by social studies of science and technology, in particular actor-network theory and Bruno Latour's (Latour 1987) analyses of 'immutable' and 'mutable mobiles,' mobilities theorists pay close attention to the infrastructures, technical objects, prostheses and embodied

Table 3 Space time combinations for platial constructions

| Time | Space | Place (S', T', meaning) | Description |
|------|-------|-------------------------|-------------|
| FT | FS | Place is driven by meaning only ("1-D") | At a specific time or fixed time interval for which only place is variable. Examples: A historical place, monument, shop, etc., which can have multiple meanings or interpretations |
| FT | CS | Driven by meaning, and changing spaces ("2-D") | On a specific time or fixed time interval, the multiple spaces defining the place. Examples: a shop extended its boundary by including neighboring empty shop with various meanings of this place |
| CT | FS | Driven by changing time and meaning ("2-D") | Space defining place is not changing over time. These are the permanent places as discussed in Table 1 |
| CT | CS | Driven by change in meaning, time and space("3-D") | Space defining the place is changing over time. The change could be change in spatial extent or displacement or both. The spatial displacements would make the case for generalized "platial mobility," while changes in spatial boundaries would lead to "dynamic places" as discussed in Table 1. An example for "platial mobility" is when communities, in case of any disaster, move from one place to another and they recreate some places like mosques, churches, community hall, etc |

*FT*, time or its interval as fixed or constant; *CT*, time or its interval as changing or variable; *FS*, fixed or constant space; *CS*, changing or variable space practices that assist or disable flows or mobility (Latour 1993; Büscher et al. 2010). Everything from shoes and bikes, mobile phones and motor vehicles, passports and satellites, software code and embedded sensors are part of the sociotechnical assemblages or human/material hybrids that perform as mobile systems and support specific mobility regimes.

One of its founder and leading proponent, Urry (2016) defines five interdependent types of mobilities as follows:

- Corporeal Travel: Travel of people for work, leisure, family life, pleasure, migration, and escape.
- Object Travel: Physical movement of objects to producers, consumers, and retailers; also sending/receiving of presents.
- Imaginative Travel: Images of places and peoples appearing in one's mind from interaction with media and associative objects.
- Virtual Travel: Travel to virtual spaces
- Communicative Travel: Travel through person-to-person messages via email, SMS, telegraph, and telephone.

Based on considerations of a diversity of travel, associated sociological considerations, as well as holding a wider notion of "society," the mobilities paradigm is a compatible sociological framework which enhances our platio-temporal representation devised earlier. In particular, we can inquire and account for how the notion of "meaning" in our representation of place is affected by social factors and influences. "Platial mobility" can therefore be viewed as a hybrid and novel concept which is a platio-temporal representation with sociological considerations, drawing from the rich mobilities sociology paradigm. Based on these ideas, we discuss their potential utility in the context of a use case which can be further revisited from our platio-temporal representation.

## 6 Platial modeling for disaster management

Natural and human-made disasters often change the meaning and the spatial extent of places over time, causing 'platial mobility.' Such an example is the movement of population, for example, internally displaced persons (IDPs) and their valuable assets and stock from their usual settlement to safe areas, in order to survive or avoid disasters. For the sake of this study, we differentiate between the terms relocation and displacement in the context of disaster risk, considering relocation as a planned and government-assisted process, with a rather voluntary character, while displacement is the abrupt and forced movement of groups of people. Relocation is a temporary or permanent (Ferris 2012) passive adaptation measure to natural hazards, defined also as the process of moving from one place to another (Pramitasari and Buchori 2018). Temporary relocation deals with the evacuation from affected areas to safe places. In this case, the platial movement is in two directions-once in the direction to the safe evacuation sites, and once back to the primary location. In the best-case scenario, subject of evacuation are humans, cattle, and valuable material objects as administrative and historical records, scientific and technical documentation, mobile cultural monuments, etc. (Zlatunova 2020). Permanent relocation from high-risk areas is the long-lasting retreat and complete abandonment of high-risk areas. Such relocation is a planned and voluntary process, taking place mostly in economically wealthy countries as it requires solid investments and the existence of a well-developed long-term disaster mitigation policies. As such, our platial mobility framework is aimed to address all stages of disaster management, namely response, recovery, adaptation and mitigation along with risk reduction. Some examples in Europe can be found in the UK (North Norfolk (The UK Coastal Change Pathfinder Programme 2012)), Germany (Simbach at the Inn (Mayr et al. 2020)), Austria (Machland, Eferding, Tirol, municipality of Goefis (Schindelegger 2018)). Temporary and permanent relocations are also common in the U.S.A. Perry and Lindell (1997), where mobile homes are a more typical way of dwelling. Although costly, planned resettlements are still considered to be the most effective passive measure for risk mitigation. Internal displacements have a rather instantaneous and long-lasting character, meaning that the disaster has caused permanent changes to the environment and it is no more inhabitable. As opposed to permanent relocation, displacement represents an unplanned and forced (involuntary) movement of people (Seebauer and Babcicky 2016; Chishtie 2018).

### 6.1 Mapping platial mobility for disaster risk reduction

Mapping platial mobility in the context of disaster management is a complex and intrinsically multidimensional task as displacements cause severe multidimensional distress, manifested at the physiological, psychological, and sociocultural levels (Correa et al. 2011). Multi-dimensionality brings along the problems and complexity of n-dimensional integration above the 3D. Multidimensional mapping requires an n-dimensional data model and database that can store n-dimensional objects and support n-dimensional data types, geometries and topological relations (Ohori et al. 2017), n-dimensional clustering and indexing (Van Oosterom and Stoter 2010).

GIS maps are essential for decision-making during crisis situations (Ginige et al. 2014) and coordination of rescue actions. Given the infusion of sociology, the platial mobility framework can improve particular application in, for example, the planning of permanent or temporary relocations. In this regard, maps derived from the platial of temporary displacements can assist evacuation planning, i.e., finding evacuation routes, allocating evacuation sites, planning of supply, analysis of the spatio-temporal movement of people and goods, etc. Maps of permanent displacements shall represent the long-term changes in land use and risk level, normally a decrease in risk level, and can be used for development of disaster management policies at different territorial levels.

In this paper, we assume a general model of risk, represented by the hazard, exposure and vulnerability dimensions. The hazard dimension carries information about the range, frequency, duration, magnitude of the disaster, thus holding spatial, temporal, scale (from individual to continental (Cutter et al. 2008)) and meaning dimensions, depending on the affected unit, while exposure is the infrastructure which are potentially affected by the disasters such as buildings, roads, and highways. The vulnerability dimension is generally represented by the human exposure to disasters and the coping capacity of the society, economy and ecosystems. Data about risk can be found in risk assessments and in the regional/national spatial data infrastructures for disaster risk management (Manfré et al. 2012) but switching from single-risk to multi-risk assessments imposes new challenges ahead the dimensional integration in data models (Babukova and Zlatunova 2016). Platial mobility models can provide rich data, particularly in the vulnerability aspect, which can then be combined with hazard and exposure layers to ascertain disaster risk in a holistic manner.

### 6.2 Mapping platial mobilities in the formation and aftermath of Lake Attabad

Lake Attabad is an artificial lake in Hunza, Gilgit-Baltistan province of Pakistan which formed on 4 January 2010 due to a landslide triggered by a huge slope failure above Attabad village (Fig. 3). This resulted in blockage of the Hunza river valley whose extent was reported at 2 km with a depth of 120 m. The village was destroyed and over the course of five months a lake formed behind the landslide site. In the aftermath, an overall five villages were affected with 240 households submerged with 380 families displaced (Cook and Butz 2013) and resided in shelters.

Those affected by this tragedy were affected heavily both from a livelihood and from the emotional and trauma viewpoints. Due to the disaster, their notion of place was disrupted and day-to-day mobility was affected heavily with the government only providing an infrequent boat service to furnish basic supplies to the displaced communities in their respective shelters (Sökefeld 2014). Cook and Butz (2013) studied the restricted mobility in the Gojal region, using traditional maps and interviews with those affected in the aftermath of the disaster. Key findings are the importance and impacts of (im)mobility on the victims as well as the social and political factors which included lack of proper disaster management strategies since the onset of the disaster. This highlights the importance of social aspects in such events and in the formation of platial representations.

While such studies are indeed valuable, we contend that such complex events caused by disasters can be studied via our considerations of platio-temporal dynamics from a GIS perspective. In this regard, we contend that mapping platial mobility via a combination of GIS and mobile methods (Büscher and Urry 2009) can be useful for sketching out both the geometries of such movements and places via simultaneously including their human and social aspects. Hence, such platio-temporal mappings will not only involve geospatial place-based locations, but via interviews, focus group discussions and following the various subjects through their everyday mobilities with various types of media will also allow the social aspects to emerge as well.

Based on the insights provided by this case, we envision further applications of our platio-framework to not only sedentary entities but also dynamical phenomena such as transportation, urbanization, public health including disease mapping, immigration, tourism and disaster management.

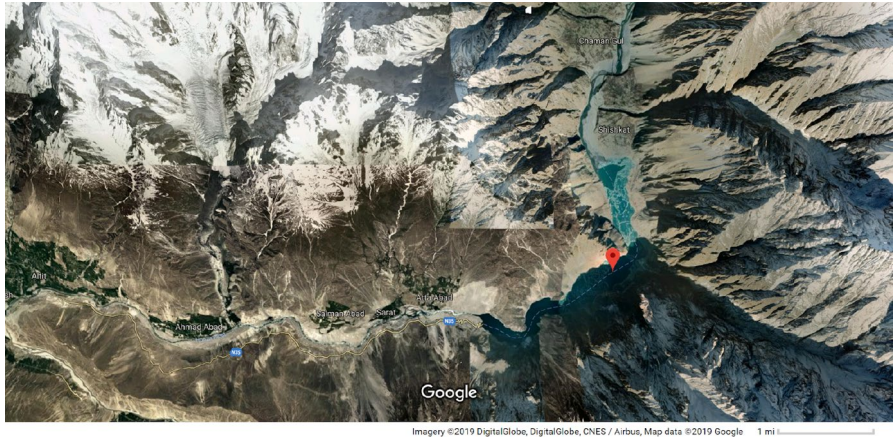

**Fig. 3** A view of Lake Attabad, courtesy of Google Maps

### 6.3 Prospects of mapping platial mobility in the Eferding basin relocation

The Eferding Basin is one of the six retention basins of the Danube river in Austria, located in the province of Upper Austria (BMLFUW 2016). The basin is an important agricultural region in Austria (Seebauer and Babcicky 2016), home of over 20 settlements and important transport corridor in the Danube plain. A significant part of the basin falls within the functional planning zones with high flood risk (probability of flooding HQ30) and middle risk (probability of flooding HQ100), defined by the federal flood risk assessment and zoning (eHORA (n.d.)). The most severe historical floods in the Eferding basin are the ones from 1899 (water discharge 8.500 m$^3$/s), 1954 (8.800 m$^3$/s), 2002 (6.600 m$^3$/s, equivalent to probability HQ25), 2013 (9.500–10.000 m$^3$/s, equivalent to probability HQ250) (see Fig. 4) (Seebauer and Babcicky 2016). The flood of June 2013 exceeded the level of the 100-year flood with 119 cm and caused direct damages to households and agriculture (Schindelegger 2018) amounting to over 253 million EUR (Anschober 2015). This event proved the inefficiency of the new flood protection measures, introduced after the flood in 2002. The flooding of the plain was controlled by the management of the water power plant Ottensheim-Wilhering. Later on, the dam management authority was criticized for not complying with the operating regulations for water release, and for taking advantage of the high electricity production during the high water levels. The local population was not warned about the controlled flooding of their land, which led to higher damages and sharp reactions to the management of the dam (ORF 2013; Kurier 2013).

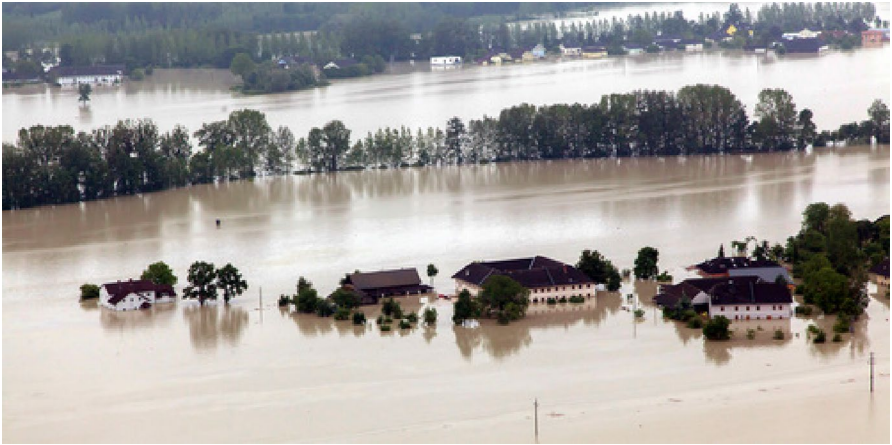

**Fig. 4** Flooded village in Eferding Basin 2013. *Image source*: police records

The flooding in 2013 was a turning point for the flood risk management policy in the region and a new budget for flood protection was allocated, among which compensation for the relocation of 146 households. The relocation zone in the Eferding basin was defined already by the end of 2013, covering an area of 24.35 km, affecting 612 buildings, 138 of which housings buildings (Seebauer and Babcicky 2016). The relocation proposal includes financial subsidy of 80 percent of the current property value, and 80 percent of the demolition costs for the abandoned property. The ownership of the abandoned lands remains unchanged but the land use type is changed from building land to grassland. By 2019, the owners of 72 properties have signed the displacement contract and 57 of them have already completely moved to the new locations.

Even when a relocation project is based on substantial planning and receives good financial assistance, people are often reluctant to move away. GIS maps that depict platial mobility of relocations can assist the decision-making process. Maps are also important tools for analysis of the mid- or long-term effects of flood risk mitigation planning policies.

Some multidimensional formulations of relocation can be found in the literature, for example, Correa et al. (2011) describe resettlement as a multidimensional process that holds (i) a physical dimension, dealing with information on real estate, private and institutional built structures, and public service infrastructure; (ii) a legal dimension that addresses land rights, relationships with public services, and the lawfulness of the settlement; (iii) a social dimension-the number of affected population and its demographics; (iv) an economic dimension-level of income of affected individuals and value of property; and also (v) cultural, (vi) psychological, (vii) environmental, (viii) political-administrative and (ix) territorial dimensions. After conducting surveys with the relocated households in Eferding, Seebauer and Winkler (2020) recognise four dimensions that influence the individual decision-making: (i) economic dimension offered financial compensation for resettlement, household income, family situation (children, retirement, multigenerational households, etc.); (ii) emotional dimension, including fear of the next natural calamity, post-disaster mental stress, emotional resilience to disasters, place belonging and identity; (iii) risk dimension, representing the risk level, uncertainty-the maintenance of social networks, relationships and (lack of) coordination among neighbors, feeling of security when neighbors decide to stay in the risk zone.

In order to apply the formalism of place for mapping disaster displacements, we suggest a four-dimensional approach, discussed in Sect. 3, that integrates platial mobility and risk dimensions. The platial mobility component is depicted by the dimensions space, time, scale and meaning. The *spatial* dimension represents the physical objects with their properties of position (location), shape (geometry, spatial dimension) in 2D or 3D, and topology. The *scale* dimension represents the platial hierarchies, also defined as a level of detail or abstraction. In the case of disaster displacements, scale can be represented by social units, i.e., household, community, business, etc. (Correa et al. 2011). The problems of generalisation can be effectively addressed by formalizing semantic relationships and using an ontology to automate the selection of algorithms (Gould et al. 2014). The latter brings in the challenges of supporting smooth zooming throughout the scales, which can be overcome with multirepresentational data models. The *temporal* dimension can represent time instances (i.e., snapshots) or time intervals within a timeline of the resettlement. Time can be modeled as a single or multiple timelines (Van Oosterom et al. 2006), depending on the level of detail (household, community, region, etc.) as owners resettle individually, at different times and pace. The timeline of a resettlement can include different instances and intervals between them, i.e., the occurrence of the disaster, the designation of relocation zones, the deadlines for signing the relocation agreements, complete removal of property from the high risk zone (Seebauer and Winkler 2020). A true integration of space, time and scale should ideally preserve the continuous topological relations between objects and events in the modeled geographic area (Van Oosterom and Stoter 2010; Arroyo Ohori et al. 2015). The *meaning* of places appears as the most problematic dimension to model and map as it stands for the emotional state, belonging to place and maintaining social relationships, which have no physical representation. While physical movement can be modeled by distance functions, changes in the meaning of places are difficult to formalize in quantitative terms. With mobile methods developed in the mobilities paradigm (Büscher et al. 2010) and our platial mobility framework, tracking the meaning of places will require qualitative analyses and methods, and some potential data sources are questionnaires, travel logs and social media data.

### 6.4 Platial mobility for studying COVID-19 and developing effective responses to pandemics

The COVID-19 pandemic is a rapidly evolving global threat with more than 6.35 million deaths to date and an ongoing rapid increase in number of cases exceeding 523 million (Bedford et al. 2020; Rothan and Byrareddy 2020; Spinelli and Pellino 2020; Velavan and Meyer 2020; Wang et al. 2020). These resulting tragedies combined with the social, health, and economic impacts of mitigation efforts to slow disease transmission such as lockdowns, mandatory self-quarantine especially after travel, social distancing, etc., are shifting human relations to limit exposure using preventive practices which also shift mobilities, call for a careful study of this present crisis from interdisciplinary and multidisciplinary aspects. In developing a deep understanding of risk and potential impacts of the COVID-19 pandemic, studying social demographics, associated mobilities and vulnerabilities, along with environmental factors are critical to developing understanding of risk factors that make a community more vulnerable. This knowledge is essential to better support currently affected communities as well as adequately prepare for future outbreaks (Djalante et al. 2020). Social and environmental factors require careful considerations in order to facilitate a better understanding of their influences involved in both epidemiological dynamics of virus transmission and related environmental impacts. Platial mobility is a suitable framework to study changes in places and mobilities using GIS.

Geospatial big data management and analytics are being used as a means to monitor and model this pandemic (Zhou et al. 2020), however, most of the data used are of epidemiological nature. Existing GIS-based platforms such as the Johns Hopkins University COVID-19 tracker (see Fig. 5) (Dong et al. 2020) and others outlined by Boulos and Geraghty (2020) are reporting such global data at various levels, including country and state levels. It is notable that these trackers are spatially coarse and while layers of street maps are available these can be further enhanced by incorporating movement of citizens. Mobility data in developed nations have been particularly useful in mapping dynamics of the pandemic (Buckee et al. 2020; Lasry et al. 2020; Lau et al. 2020; Pepe et al. 2020). These mobility studies using movement of traffic and populations are done in various cities such as San Francisco, USA and across Italy afflicted by the pandemic. Using de-identified and aggregated mobility data via mobile positioning technology, which is both spatially and temporally aggregated, studying patterns of movement can potentially provide insights into social distancing and changed movement patterns across scales. By including these mobility datasets along with an epidemiological tracker, our platial mobility framework can help enhance these present studies to better understand social and spatio-temporal dynamics by mapping shifting notions of places, mobilities and vulnerabilities using geospatial technologies as well as collection and integration of qualitative data. This is expected to result in localized information which can be collected via online forms and also via virtual call-ins with consenting participants who can describe their day-to-day mobilities. Collecting this information and mapping them along with existing epidemiological data will help display and provide patterns of study for shifting places and movements for participants experiencing lockdowns and social distancing measures. Design of specific web-based dashboards on localities and cities, especially for Lower and Middle Income Countries (LMICs), is also great opportunities to meet the data gaps here for increasing awareness evidence-based decision-making. Visualization of these movements can potentially be offered to the public to help display clustering behavior in particular localities, along with reasons provided in qualitative data collection for researchers to study as to how and why those affected are shifting their daily movements. With platial mobility mapping, we can therefore create enhanced understandings of complex and shifting movements from both a GIS and sociological points of view, which can further help develop better responses to this ongoing COVID-19 crises and also help address upcoming epidemics and pandemics in the days ahead.

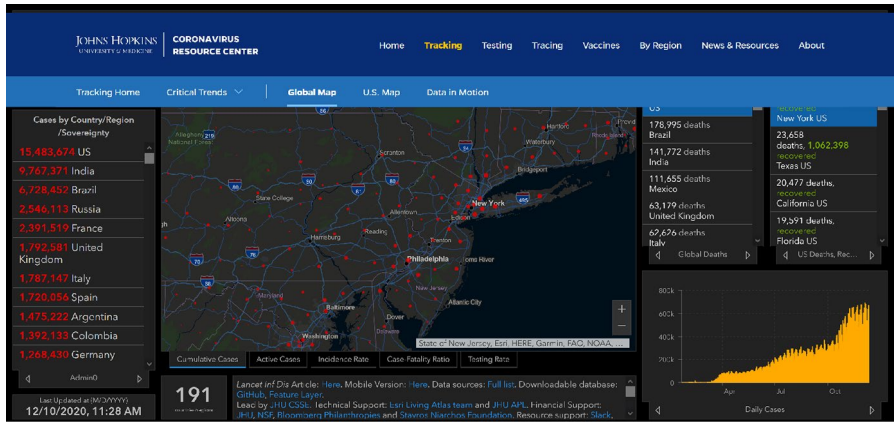

**Fig. 5** The Johns Hopkins COVID-19 Web-based data Dashboard

## 7 Conclusions

In this work, we have developed a novel conceptual framework denoted as "platial mobility," with considerations of space, time, and meaning integrated with sociology via the mobilities paradigm in order to faithfully represent both sedentary and dynamical entities for better mapping and related applications.

Our platial formalism is based on the notion of place that is time dependent and does not consider 'space' and 'place' to be necessarily separated. So places are semantically constructed spaces or sets of spaces that have spatial (geometry and topology), temporal (time) and meaning (semantics) dimensions. We discuss three important aspects of platial classification, platial construction and platial comparisons. We provide a classification of places and their salient characteristics and places can be temporary or permanent, movable or immovable, and instantaneous or planned.

Places may be constructed hierarchically, in that case an additional dimension of level of detail is considered. Places may have hierarchical (place/sub-place) relationships that are valuable in performing platial analysis. The hierarchies are defined along spatial and semantic dimensions thus enabling the level of detail based representation of places. Since places are semantically constructed spaces by individuals or groups, that may have sense in three different scales arranged hierarchically from local to global levels. Places may have crisp or fuzzy and imprecise boundaries defined by the spatial dimension. Although approaches exist for handling the fuzziness issue in the spatial component, fuzziness may also exist along meaning dimensions that makes the platial modeling task even more challenging. Places can be compared for similarity using some distance measure along spatial, temporal and meaning dimensions. Platial semantic similarity is the similarity along meaning dimensions and can have different contexts based on the application domain.

We list four requirements that a platial model should fulfill in order to be practical. In view of these requirements, our formalism approach defines place as a function dependent on the key dimensions of space, time, meaning and an optional dimension of the level of detail. Meaning is the most distinguishing component of our platial definition and is multidimensional carrying the subjective semantic information associated with the place. We then differentiate the spatial and platial models along level of detail, time, space (geometry and topology) and the meaning dimensions. We define the variability of place as the divergence of two places in meaning that can then be formulated in terms of their spatial and temporal dimensions. In this context, we introduce various possible platial constructions based on space time variations.

With these considerations, we introduce the idea of 'platial mobility' that draws upon the mobilities paradigm that is a contemporary framework in social sciences and studies the movement of people, ideas, things/objects and societal consequences of these flows. We therefore expand existing sedentary notions of place-based GIS and also extend mobility in GIS using the mobilities paradigm in our framework. We consider this to have a paradigm shift level impact as GISs will incorporate new types of data and change in the basis of state of science incorporating multi-dimensional data to addressing common to highly complex problems.

To highlight some areas in which "platial mobility" can contribute significantly, we have presented two cases involving improvement of disaster management and how disaster risk reduction can benefit from our platio-temporal representations. We also address the present and future epidemics and pandemics, and rapidly evolving COVID-19 pandemic, which can be studied and understood via our framework. These use cases highlight both the potential and relevance of our framework, especially in the context of ongoing and future climate change impacts and pandemics.

For our aforementioned use cases, we intend to apply and test our framework empirically, with upcoming research studies. Based on their context, we aim to use case studies and mixed-methods based approaches as our methodological basis. Overall, our aim is to improve capabilities of GISs toward meeting demands and enhancing capacity of users, for the widespread benefit to communities of disciplinary practices to beneficiaries, including vulnerable communities and general public.

**Funding** Open access funding provided by Graz University of Technology.

**Publisher's Note** Springer Nature remains neutral with regard to jurisdictional claims in published maps and institutional affiliations.

## Authors and Affiliations

**Farrukh Chishtie[1,2]**  **· Rizwan Bulbul[3] · Panka Babukova[3] · Johannes Scholz[3]**

    Farrukh Chishtie
    fachisht@uwo.ca

    Panka Babukova
    babukova@student.tugraz.at

    Johannes Scholz
    johannes.scholz@tugraz.at

[1] Department of Occupational Science & Occupational Therapy, University of British Columbia, Vancouver, BC, Canada

[2] Peaceful Society, Science and Innovation Foundation, Vancouver, BC, Canada

[3] Research Group Geoinformation, Institute of Geodesy, Graz University of Technology, 8010 Graz, Austria